\documentclass[sigconf,authorversion,nonacm]{acmart}

\usepackage{braket}
\usepackage{amsmath}
\usepackage{subfigure}

\AtBeginDocument{%
  }

\begin{document}

\title{
SPulseGen: Succinct pulse generator architecture maximizing gate fidelity for superconducting quantum computers} 

\author{Ryosuke Matsuo}
\affiliation{%
  \institution{Osaka University}
  \country{Japan}}
\email{matsuo.ryosuke.ist@osaka-u.ac.jp}

\author{Kazuhisa Ogawa}
\affiliation{%
  \institution{QIQB Osaka Univ.}
  \country{Japan}}
\email{k-ogawa.qiqb@osaka-u.ac.jp}

\author{Hidehisa Shiomi}
\affiliation{%
  \institution{QIQB Osaka Univ.}
  \institution{QuEL, Inc.}
  \country{Japan}}
\email{shiomi.hidehisa.qiqb@osaka-u.ac.jp}

\author{Makoto Negoro}
\affiliation{%
  \institution{QIQB Osaka Univ.}
  \institution{QuEL, Inc.}
  \country{Japan}}
\email{negoro.makoto.qiqb@osaka-u.ac.jp}

\author{Takefumi Miyoshi}
\affiliation{%
  \institution{QuEL, Inc.}
  \institution{e-trees.Japan, Inc.}
  \institution{QIQB Osaka Univ.}
  \country{Japan}}
\email{miyoshi@quel-inc.com}

\author{Michihiro Shintani}
\affiliation{%
  \institution{Kyoto Institute of Technology}
  \country{Japan}}
\email{shintani@kit.ac.jp}

\author{Hiromitsu Awano}
\affiliation{%
  \institution{Kyoto University}
  \country{Japan}}
\email{awano@i.kyoto-u.ac.jp}

\author{Takashi Sato}
\affiliation{%
  \institution{Kyoto University}
  \country{Japan}}
\email{takashi@i.kyoto-u.ac.jp}

\author{Jun Shiomi}
\affiliation{%
  \institution{Osaka University}
  \country{Japan}}
\email{shiomi-jun.ist@osaka-u.ac.jp}
\renewcommand{\shortauthors}{Name et al.}

\begin{abstract}
This paper proposes a cost-effective architecture for an RF pulse generator for superconducting qubits.
Most existing works use arbitrary waveform generators (AWGs)
that require both a large amount of high-bandwidth memories and high-performance analog circuits to achieve the highest 
gate fidelity with an optimized RF pulse waveform.
The proposed pulse generator architecture significantly simplifies both the generator circuit and the waveform of the RF pulse to a cost-aware square pulses.
This architecture eliminates the requirement for power- and cost-intensive AWG, a major obstacle in realizing scalable quantum computers.
Additionally, this paper proposes a process to optimize pulse waveforms to maximize fidelity of gate operations for single and multiple qubits.
Quantum dynamics simulation of transmon qubits, wherein the state of system evolves with time, demonstrates that our pulse generator 
can achieve practically the same gate fidelity as ideal RF pulses, while substantially reducing the performance requirements of memory and analog circuits.
\end{abstract}





\maketitle

\section{Introduction}
Quantum computers have attracted increasing interest in recent years, as they are considered to surpass the computational power of ``classical'' computers in certain domains. 
Various qubits have been demonstrated, such as with trapped ions~\cite{ion-trap} or photonic qubits~\cite{photonic-quantum}.
Among others, the superconducting transmon qubit is a promising candidate to realize large-scale quantum computing systems~\cite{Wendin_2017}. 
A state-of-the-art superconducting quantum computer features 433 transmon qubits~\cite{gambetta2020ibm}.
Enhancing the number of qubits while suppressing the error rate helps to exponentially enhance the quantum computer's performance, making it the crucial guiding principle for achieving quantum supremacy~\cite{surface-code}.

In current implementations of superconducting quantum computers, gate operations for qubits are executed with the help of classical computers. 
This necessitates a strong need for control systems to be scalable with the pace of qubits. 
The transmon qubits in a quantum computer use different oscillation frequencies to avoid unnecessary coupling with neighbor qubits~\cite{qubit_quide}.
In addition, the oscillation frequency is prone to manufacturing variation.
Therefore, customized RF pulses must be designed and applied for the quantum gate operations of each qubit. 
These circuits, generating RF pulses for quantum gate operations, are referred to as qubit controllers.


Well-known quantum algorithms, such as Shor's algorithm, require at least $10^5$ to $10^6$ qubits \cite{surface-code} to execute.
Accordingly, qubit controllers must also be scalable. 
Presently, research on qubit controllers relies on high-speed arbitrary waveform generators (AWGs) to generate elaborated waveforms for maximizing gate fidelity, often disregarding circuit costs.
In general, an AWG is comprised of high-bandwidth memories (HBMs) and a fast digital-to-analog converter (DAC).
AWGs are an obstacle to scalability due to their high-performance DACs and large memory capacity.


To improve the scalability of qubit control systems, cryo-CMOS qubit controllers are proposed \cite{cryo-controller-ISSCC2020, cryo-controller-ISSCC2022, cryo-controller-VLSI2021, cryo-controller-ISSCC2019, cryo-controller-ISSCC2021}.
Providing a control line for every qubit in the 10~mK environment from room temperature is not feasible for a $10^6$ qubit system due to multiple factors, such as RF loss, mechanical congestion, heat load, and connector unreliability.
One of the design objectives of cryo-CMOS qubit controllers is to reduce both power consumption and area since the cooling power of the cryogenic refrigerator (e.g., 4~K) is limited to a few watts.
However, these cryo-CMOS qubit controllers still rely on AWGs for generating intricate RF pulses.

Therefore, this paper proposes a cost-effective qubit controller architecture that eliminates AWG.
This paper also presents a tuning method to maximize gate fidelity of RF pulses generated by our architecture.

The contributions of this paper are summarized as follows.
\begin{itemize}
    \item This paper proposes a succinct RF pulse generator architecture that replaces the conventional AWG for scalable quantum computing.
    Our architecture relaxes the memory capacity and data rate requirements, by two orders of magnitude compared to existing qubit controllers. 
    \item  
    A design methodology is proposed for optimizing pulse shape under the proposed novel architecture while maintaining maximum gate fidelity.  
    To the best of our knowledge, this paper is the first attempt to optimize RF pulse 
    considering gate fidelity.
    \item Dynamics simulation of transmons using QuTiP~\cite{Qutip}, an efficient and accurate quantum systems simulator verified with various measurements, demonstrates that gate fidelity of our pulse generator is comparable to the best AWG-based qubit controllers
    in randomized benchmarking.
\end{itemize}

The rest of this paper is organized as follows.
Section~\ref{sec:preliminary} describes transmon qubits and existing AWG-based qubit controllers.
Section~\ref{sec:proposed-controller} proposes cost-effective qubit controller architecture.
Section~\ref{sec:pulse-optimization} discusses a methodology for maximizing gate fidelity of RF pulses and then simulation results demonstrate our approach can reduce circuit costs without compromising gate fidelity.
Finally, Section~\ref{sec:conclusion} concludes this paper.

\section{Preliminary}
\label{sec:preliminary}
This section provides a brief overview of transmon qubits and the quantum gate operations with RF pulses. Please refer to~\cite{qubit_quide} for a comprehensive description on qubit control.

\subsection{Transmon qubits}\label{sec:2.2}
A transmon qubit is an anharmonic oscillator, in which
the lowest two energy levels $\ket{0}$ (ground state) and $\ket{1}$ (excited state) form a computational subspace.
The quantum state is represented by $\ket{\psi} = \alpha\ket{0}+\beta\ket{1}$, where $\alpha$ and $\beta$ are complex probability amplitudes and satisfy $|\alpha|^2+|\beta|^2 = 1$.
The quantum state may be lost due to the following two effects.
The first effect is energy relaxation and dephasing. The quantum state decays nearly exponentially with time $T_1$ and $T_2$, each corresponding to the energy relaxation and dephasing, respectively.
By the energy relaxation, a qubit in the excited state $\ket{1}$ loses energy and converges into the ground state $\ket{0}$ over time.
By dephasing, the relative phase of the quantum state is lost over time.
The second effect is a transition to non-computational states. Transmon qubits can take other energy levels than $\ket{0}$ and $\ket{1}$.
Such energy levels are called non-computational states.
This happens when a transmon receives an RF pulse at which
the state oscillates between the computational and non-computational states.
In general, we only consider the third lowest energy level $\ket{2}$ as a non-computational state since the transition to higher energy levels than $\ket{2}$ is negligible.
In this paper, the transition frequency
where the energy state oscillates between $\ket{0}$ and $\ket{1}$
is denoted as $\omega_{0\rightarrow1}$, while that with
the non-computational state is denoted as $\omega_{1\rightarrow2}$.
In transmon qubits, the transition frequencies $\omega_{0\rightarrow1}$ and $\omega_{1\rightarrow2}$ are designed to be sufficiently different, which allows us to individually address these two transitions.
In this paper, a qubit frequency $\omega_q$ indicates $\omega_{0\rightarrow1}$ of the qubit.
The anharmonicity $\alpha = \omega_{1\rightarrow2} - \omega_{0\rightarrow1}$ is negative and can not be made arbitrarily large.

\subsection{Quantum gates using RF pulses}
\subsubsection{RF pulse for universal gate}
By adjusting the amplitude and phase of the RF pulse
fed into transmon qubits,
multiple quantum gate operations can be implemented.
In transmon quantum computers, the gate fidelity represents the accuracy of gate operation via RF pulses. 
The three operations, "Virtual Z gate," "derivative reduction by adiabatic gate (DRAG)," and "Cross resonance (CR) gate,"  are applied in specific sequences to construct universal gate sets.
The virtual Z gate does not require an individual RF pulse because the Z operation can be embedded into the other gates by adding a corresponding phase.

\subsubsection{DRAG}
The DRAG pulse drives a qubit using the RF pulse formulated by Eq.~(\ref{eq:DRAG}).
\begin{equation}
\label{eq:DRAG}
    v(t) = I(t)\cos(\omega_dt + \phi) + Q(t)\sin(\omega_dt + \phi).
\end{equation}
Here, $v(t)$ is a transient voltage of the DRAG pulse. 
$I(t)$ and $Q(t)$ denote in-phase and quadrature pulses, respectively.
$\omega_d$ denotes the frequency of a driving pulse, and
$\phi$ is the RF signal phase.
When $\omega_d = \omega_q$ and $\phi=0$ hold, an in-phase pulse results in rotations around the $x$-axis, while a quadrature pulse corresponds to the rotation around the $y$-axis on the Bloch sphere.
The rotation angles around the $x$-axis and the $y$-axis are determined by $I(t)$ and $Q(t)$, respectively.
It is important for $\omega_d$ to exactly match $\omega_q$ in order to avoid discrepancies
between actual and ideal rotations.
The rotation axis can be controlled by adjusting $\phi$, enabling to implement arbitrary single-qubit gate operations using single DRAG pulses.

\subsubsection{CR gate}
The CR gate pulse drives a control qubit using the RF pulse with the target qubit frequency.
When the control qubit state is $\ket{1}$, the phase of the target qubit rotation changes depending on the amplitude and duration of the RF pulse.
When the control qubit state is $\ket{0}$, the phase of the target qubit rotation does not change.
We can implement a two-qubit gate operation by exploiting these phenomena of the target qubit.

\subsection{Qubit controller}
\begin{figure}[t]
    \centering
    \includegraphics[clip, width=1.0\columnwidth]{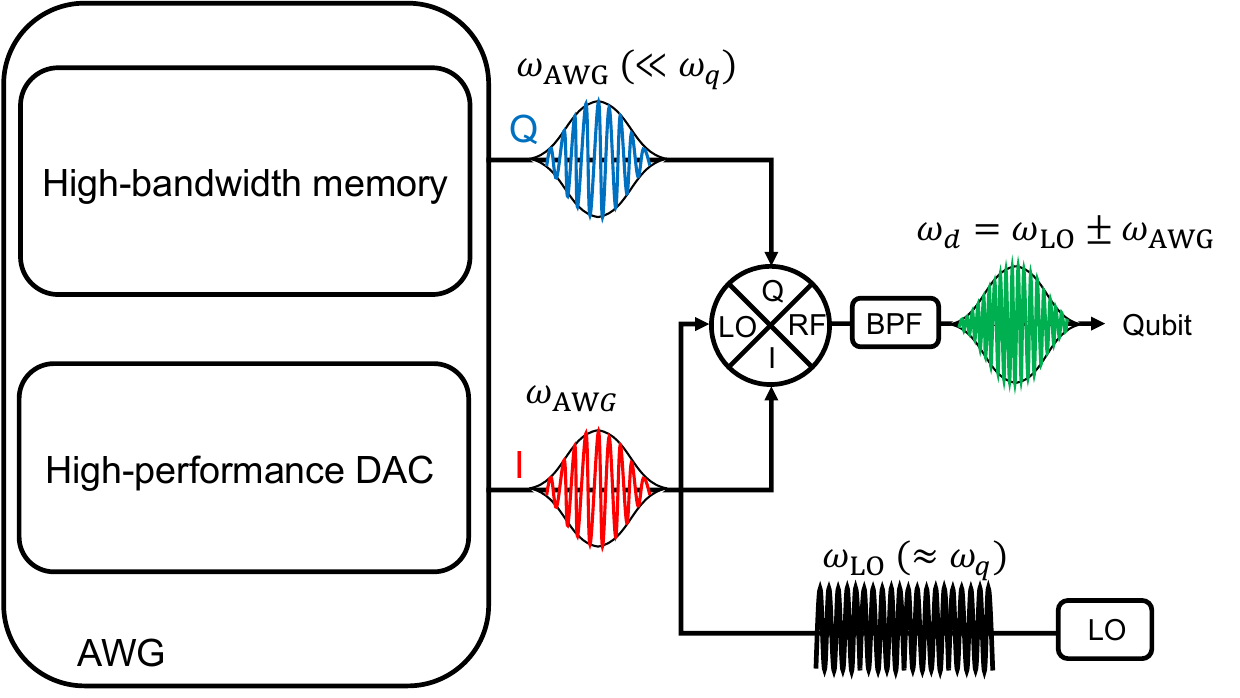}
    \caption{Schematic of AWG-based Qubit controller.}
    \label{fig:existing-controller}
\end{figure}

Fig.~\ref{fig:existing-controller} shows a schematic outline of existing qubit controller~\cite{qubit_quide}.
A local oscillator (LO) generates a carrier waveform with a fixed frequency $\omega_{\text{LO}}$ that is close to the qubit frequency $\omega_q$.
The LO's output signal is shared with multiple qubits.
The qubit frequency ($\omega_q$) differs by qubits.
$\omega_q$ also fluctuates over time.
The AWG's output signal includes a carrier frequency $\omega_{\text{AWG}}$ which is much smaller than $\omega_{\text{LO}}$ for compensating the variation.
The AWG's output signal also includes a specific envelope corresponding to each quantum gate operation.
The AWG's output signal then goes through a mixer with the LO's output signal, yielding
an RF pulse with the envelope and frequency $\omega_d = \omega_{\text{LO}} + \omega_{\text{AWG}}$. Through a bandpass filter (BPF), the pulse is fed to control a qubit.
In general, $\omega_{\text{AWG}}$ is hundreds of MHz.
On the other hand, $\omega_{\text{LO}}$ is several GHz.
The existing qubit controllers maximize the gate fidelity by generating precise
RF waveforms with AWGs.
This drastically increases the power and the cost.
This paper thus proposes AWG-less succinct architecture. 

An AWG requires HBMs to store a pulse waveform and a high-performance DAC to generate a pulse.
For example, when a pulse duration is 100~ns and
a sampling rate is 1~GS/s, at least 100 sampling points must be stored in the HBM.
To achieve scalability, the development of
a cost-effective qubit controller that
substitutes the AWG with a compact memory and
low-cost envelope generator is crucial.


\section{Cost-Effective Qubit Controller}
\label{sec:proposed-controller}

\subsection{Proposed architecture (SPluseGen)}
\begin{figure}[t]
    \centering
    \includegraphics[clip, width = 1.0\columnwidth]{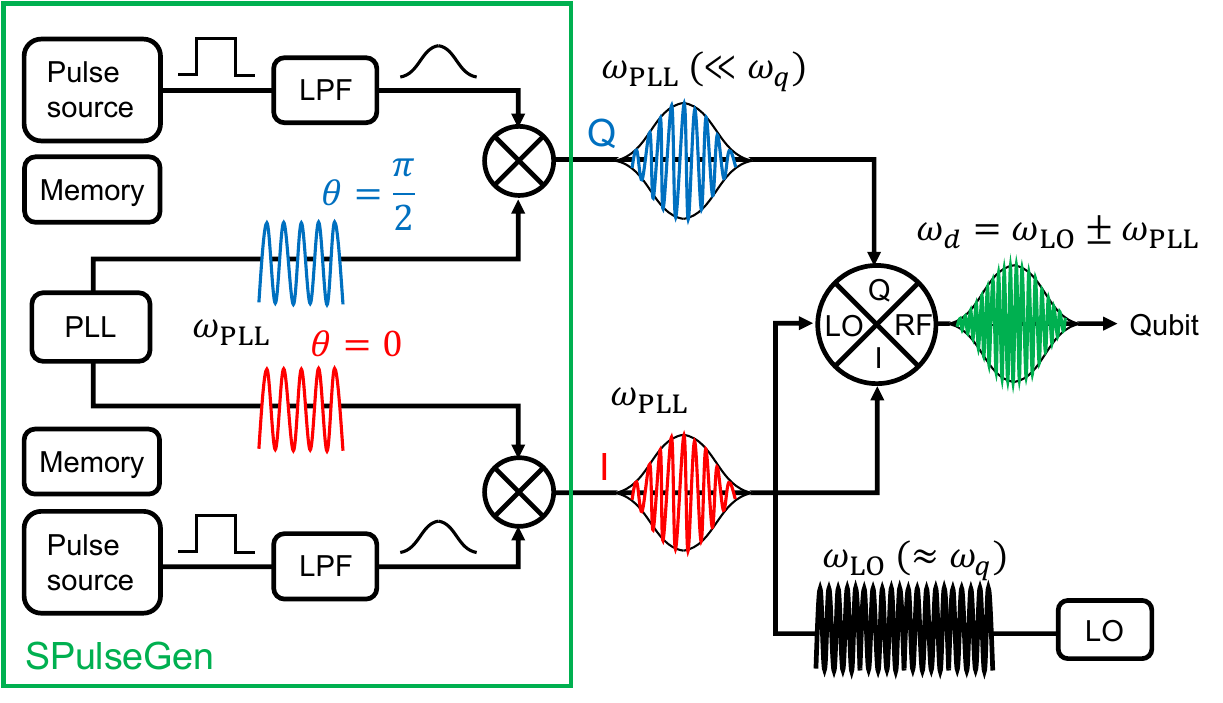}
    \caption{Proposed RF pulse generator architecture.}
    \label{fig:proposed-controller}
\end{figure}

Fig.~\ref{fig:proposed-controller} shows the proposed
architecture (SPulseGen), comprising a phase-locked loop (PLL), pulse sources, and memories.
We can adjust $\omega_d$ to $\omega_q$ by tuning carrier wave frequency $\omega_{\text{PLL}}$ generated by a PLL. 
The pulse sources generate a pulse envelope, and their low-frequency components are extracted using respective low-pass filters (LPFs). Subsequently, these pulses are modulated with carrier waves produced by the PLL with phase angles
$\theta = 0$ and $\theta = \frac{\pi}{2}$ to generate
in-phase and quadrature pulses, respectively.
The width and amplitude of the envelope pulses are adjusted for each qubit operation. 
A detailed discussion on optimizing envelope pulses will be provided in Section~\ref{sec:pulse-optimization}.

Our architecture can leverage a pulse source with a slower sampling rate and a commodity memory with a much smaller capacity compared to AWG-based qubit controllers since a free-running PLL directly generates the carrier wave with frequency $\omega_{\text{PLL}}$.
This represents a fundamental advantage of our architecture. In contrast,
existing qubit controllers generate a pulse with frequency $\omega_{\text{AWG}}$ and an envelope solely using an AWG, necessitating a high-speed DAC and a large memory.
In other words, the waveform generation in the proposed architectures is akin to drawing the shape of an amplitude-changing sinusoidal waveform using a simple filter with a continuous oscillation source, while the conventional architecture draws the waveform dot-by-dot.

\begin{figure}[t]
    \centering
    \includegraphics[clip, width=0.9\columnwidth]{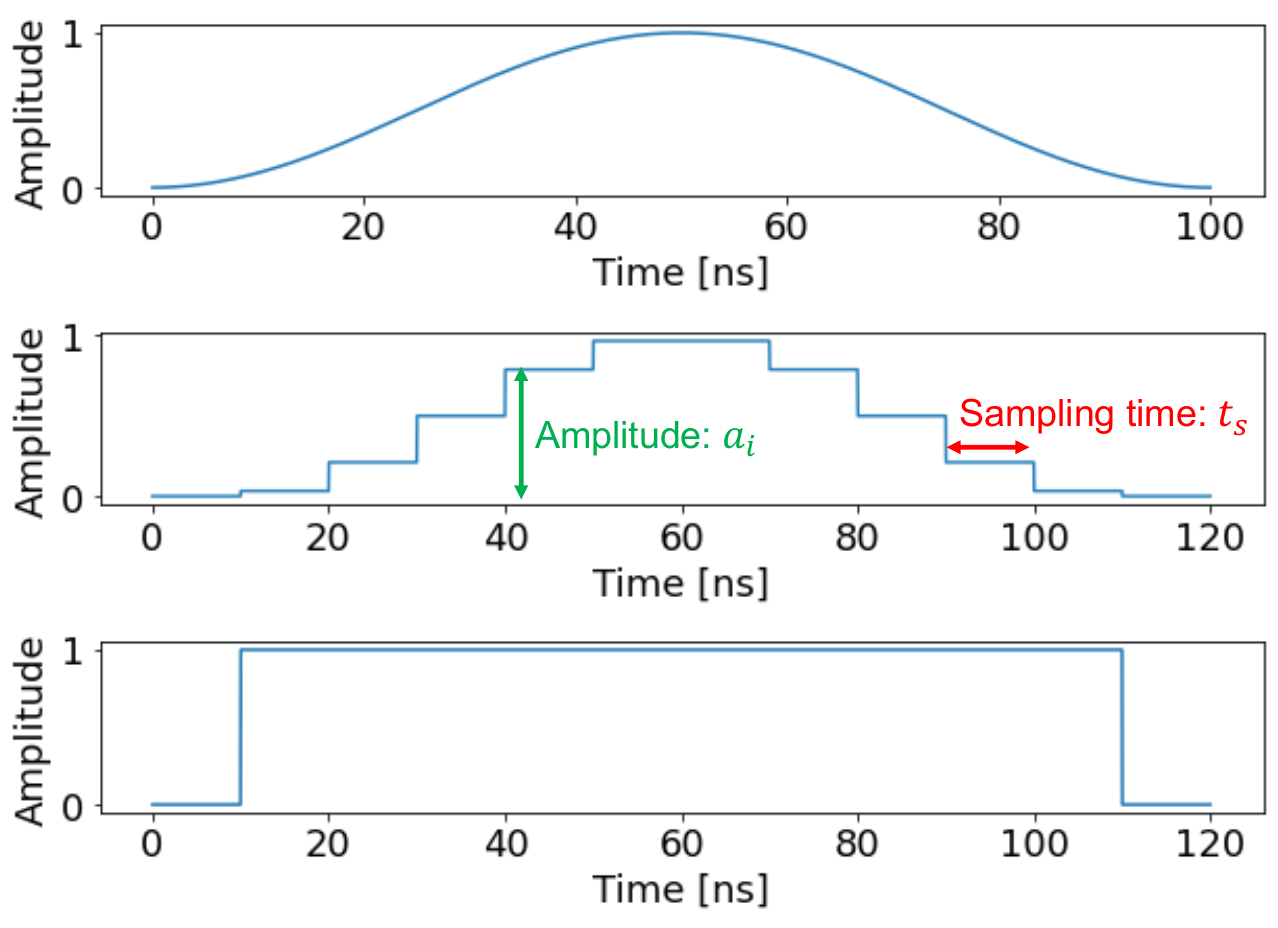}
    \caption{Envelope waveforms. Top: Ideal waveform. Middle: Staircase waveform (low-cost AWG mode). Bottom: Square waveform (AWG-less mode).}
    \label{fig:envelope}
\end{figure}

Simplifying an envelope waveform leads to further reductions in the sampling rate of a pulse source and waveform memory capacity.
In optimizing the envelope,
we consider three types of envelope waveforms shown in Fig.~\ref{fig:envelope}: (1) ideal, (2) staircase (low-cost AWG mode), and (3) square (AWG-less mode).
Reproducing an ideal waveform requires an extremely fast sampling rate and a large memory capacity to generate a smooth waveform.
A staircase waveform with a sampling speed $t_s$ (low-cost AWG mode) requires a pulse source with a sampling rate of $f_s = 1 / t_s$, reducing
the required waveform memory capacity proportional to $f_s$.
If $f_s$ is decreased, the DAC performance is also relaxed,
leading to lower design costs and power consumption.
This paper demonstrates using an $f_s$ 4.8 times smaller 
than that of state-of-the-art products still yields a comparable gate fidelity.
In the case of the square waveform (AWG-less mode), a pulse source generates a single amplitude, and a waveform memory is eventually eliminated as the amplitude is adjusted by the supply voltage.
Therefore, adopting a square waveform for an RF pulse envelope drastically reduces circuit costs.

However, employing these simpler waveforms for RF pulse envelopes may potentially deteriorate gate fidelity.
This paper quantitatively evaluates factors influencing gate fidelity with different envelopes and subsequently introduces a pulse-tuning method to maximize gate fidelity, taking these factors into account.

\subsection{Gate fidelity assessment}
\label{sec:fidelity}
The rotation of the qubit(s) in a Bloch sphere can be achieved with a series of pulses having an appropriate combination of frequency, amplitude, and duration. However,
gate fidelity may still be compromised by two factors: unwanted transitions and qubit decay~\cite{qubit_quide, off-resonance-error}.
Practical RF pulses unavoidably contain frequency components other than the desired one.
These frequency components cause unwanted transitions of a qubit state, which degrades gate fidelity~\cite{qubit_quide, off-resonance-error}.
Therefore, Sections~\ref{sec:DRAG} and \ref{sec:CR} respectively address pulse tuning methods for suppressing undesirable transitions for the DRAG pulse and the CR gate.
The aim of these methods is to achieve comparable fidelity to the ideal RF pulse generated by AWGs with sufficiently high-speed sampling rates and sufficient memory capacity.

The qubit state exponentially decays with time due to the energy relaxation and echo dephasing \cite{qubit_quide}.
A long pulse duration increases the effect of the energy relaxation and echo dephasing, which reduces gate fidelity.
Therefore, Section~\ref{sec:plen} proposes a pulse tuning methods reducing the effect of qubit decay.


\section{Pulse Envelope Optimization and Experimental Results}
\label{sec:pulse-optimization}
This section proposes tuning methods for staircase waveforms and square waveforms that prevent both undesired transitions and qubit decay.
Simulations demonstrate that our method can achieve comparable gate fidelity to ideal waveforms.
In our analysis, gate fidelity is obtained using randomized benchmarking (RB)~\cite{RB}.
Due to the statistical nature of RB,
the mean and standard deviation of gate fidelity are calculated.
Our simulation uses QuTiP \cite{Qutip}, an open-source software for simulating quantum systems dynamics.
We assume a two-qubit system ($q_1$ and $q_2$) with the following parameters: $\omega_{q_1} / 2\pi = 7500$~MHz, $\alpha_{q_1} / 2\pi = -380$~MHz, $\omega_{q_2} / 2\pi = 8500$~MHz, $\alpha_{q_2} / 2\pi = -420$~MHz, $T_1 = 260$~$\mu$s and $T_2=170$~$\mu$s \cite{qubit-param1, qubit-param2}.
Since DRAG and CR gates can represent any gate operations in a larger system, the analysis of the two-qubit system is necessary and sufficient.

\subsection{DRAG}
\label{sec:DRAG}
We assume the waveform in Eq.~(\ref{eq:ideal-DRAG}) as an ideal waveform.
\begin{equation}
\label{eq:ideal-DRAG}
    v(t) = A\cdot \bigg[1 - \cos\bigg(\frac{2\pi t}{T_p}\bigg)\bigg]\cos(\omega_dt + \phi) + B \cdot \sin\bigg(\frac{2\pi t}{T_p}\bigg)\cos(\omega_dt + \phi)
\end{equation}
where $A, B$ denote the amplitudes, and $T_p$ denotes the pulse length.
This waveform is one of the best choices for a DRAG pulse waveform.
Finding the truly optimal DRAG pulse waveform is an open problem ~\cite{qubit_quide}.
We found through simulation the best amplitude values for $A, B$ to realize $\pi/2$ rotation by sweeping amplitudes while fixing $T_p = 20$~ns.
For $q_1$, the optimal amplitudes $A$ and $B$ were $25.04$~MHz and $-1.72$~MHz, respectively.
For $q_2$, the optimal amplitudes $A$ and $B$ were $25.06$~MHz and $-1.55$~MHz, respectively.
The best amplitudes for $q_1$ and $q_2$ are different due to the differences in qubit frequency and anharmonicity.

Note that this paper uses frequency as the unit of amplitude.
Applying an RF pulse to a qubit induces rotation on the Bloch sphere, known as Rabi oscillation.
In the research field of qubit control, the amplitude is represented by the value corresponding to the frequency of Rabi oscillation.
For example, when the frequency of Rabi oscillation is 1, the amplitude of the RF pulse is represented by 1~Hz.

\begin{table}[t]
    \centering
    \caption{Expectation values of non-computational state when a DRAG pulse is applied to qubits $q_1$ and $q_2$.}
    \begin{tabular}{c|c|c} \hline
         &  Ideal & Square \\ \hline
        $q_1$ & $4.6 \times 10^{-7}$ & $5.4 \times 10^{-4}$ \\ \hline
        $q_2$ &  $1.0 \times 10^{-6}$ & $4.5 \times 10^{-4}$ \\ \hline
    \end{tabular}
    \label{tab:exp-leak-drag}
\end{table}

We aim to achieve comparable gate fidelity to the ideal RF pulse by using simpler waveforms, such as a square waveform.
The pulse length is set to 20~ns, matching that of the ideal RF pulse.
We adjust the amplitude to match the area under the envelope to that of the ideal RF pulse
since the rotation angle of the DRAG pulse is determined by the RF pulse envelope area.

In the DRAG pulse, we regard the non-computational state as an unwanted transition.
An RF pulse with frequency components around $\omega_q - \alpha$, the qubit state can transit to a non-computational state $\ket{2}$.
To evaluate the likeliness of this unwanted transition, we simulate and compare the expectation value of $\ket{2}$ when the ideal or the square waveform pulse is applied to the qubits.
The results are summarized in Table~\ref{tab:exp-leak-drag}.
Utilizing square waveform increases the transition to $\ket{2}$ state, which possibly reduces the gate fidelity.
To evaluate the gate fidelity deterioration, we simulated gate fidelity of single-qubit gate operations using the ideal waveform or the square waveform.
Infidelity denotes ($1-$ fidelity). 
The results in Table~\ref{tab:fidelity-single} demonstrate that gate fidelity of the square waveform is smaller than that of the ideal waveform.
However, this reduction in gate fidelity is considered negligibly small when considering other sources of implementation error, such as the RF pulse jitter, which causes off-axis rotations~\cite {implementation-error}.



\begin{table}[t]
    \centering
    \caption{Randomized benchmarking results of single-qubit gates.}
    \begin{tabular}{c|c|c} \hline
        & Ideal & Square\\ \hline
        Infidelity [\%] & $0.096 \pm 0.026$ & $0.173 \pm 0.015$ \\ \hline
    \end{tabular}
    \label{tab:fidelity-single}
\end{table}

\subsection{CR gate}
\label{sec:CR}
We assume the raised cosine flattop waveform formulated by Eq.~(\ref{eq:ideal-CR}) as the ideal waveform.
\begin{equation}
\label{eq:ideal-CR}
    v(t) = 
    \begin{cases}
        \frac{A}{2} \left(1 - \cos\left(\frac{\pi t}{T_r} \right) \right) & (0 \leq t< T_r) \\
        A & (T_r \leq t < T_r + T_f) \\
        \frac{A}{2} \left(1 - \cos\left(\frac{\pi (T_r + T_f - t)}{T_r} \right) \right) & (T_r + T_f \leq t < 2T_r + T_f) 
    \end{cases}
\end{equation}
where $T_r$ and $T_f$ are a rise time and a duration of flat regime, respectively.
This waveform is a well-known CR gate pulse waveform achieving a sufficiently good gate fidelity although
the exact optimal CR gate pulse waveform is an open problem \cite{off-resonance-error}.
We denote $q_1$ and $q_2$ as the control qubit and the target qubit, respectively.
$T_f$ in (\ref{eq:ideal-CR})
is optimized by 
sweeping $T_f$ under fixed
parameters of $T_r = 50$~ns and $A=300$~MHz.
The optimal $T_f$ changing the phase of target qubit rotation by $\pi/2$ is $96$~ns. 
In this paper, the optimal CR gate denotes the CR gate realizing this operation. 
Combining this CR gate and DRAG can implement arbitrary two-qubit gate operations.
In other words, the optimal pulse length $T_p$ is $2T_r + T_f=196$~ns.

In CR gates, we consider oscillations between $\ket{0}$ and $\ket{1}$ of the control qubit as unwanted transitions.
The CR gate drives a control qubit at the target qubit frequency.
If an RF pulse has frequency components close to that of the control qubit, 
an oscillation between $\ket{0}$ and $\ket{1}$ of the control qubit is initiated.
Since the CR gate must not change the state of the control qubit, this oscillation reduces the gate fidelity.
To evaluate this effect, we simulate the expectation value of Z-measurement for the control qubit when 
different RF pulses are applied to a CR gate.
The expectation value of Z-measurement is a continuous value between $-1$ and $+1$, representing a qubit state $\psi$. 
For example, the value of $+1$($-1$) indicates $\psi = \ket{0}(\ket{1})$.
Therefore, the difference in the expectation value of Z-measurement between the initial state and the state after applying CR gate, which we define as Z-error, represents the transition of the qubit state by an oscillation between $\ket{0}$ and $\ket{1}$ of the control qubit.
We now optimize the staircases and the square waveform pulses to reduce Z-errors.

\subsubsection{Staircase waveform (low-cost AWG mode)}
We consider optimizing the amplitudes series, $a_0, a_1,...$ of a staircase waveform with sampling time $t_s$ shown in Fig.~\ref{fig:envelope}~(Middle).
The sampling rate $f_s$ of a pulse source equals $1/t_s$.
We determine the amplitude $a_i$ so that $a_i \times t_s$ equals the area of the ideal waveform from $i\times t_s$ to $(i+1)\times t_s$.

\begin{figure}[b]
    \centering
    \includegraphics[clip, width=0.9\columnwidth]{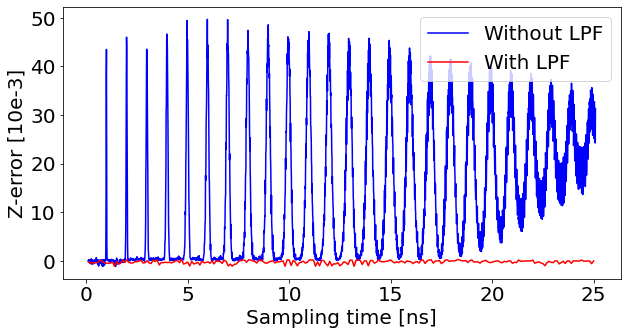}
    \caption{Z-errors obtained by varying $t_s$ in a staircase waveform pulse.}
    \label{fig:samp-rabi}
\end{figure}

The sampling time $t_s$ determines the frequency spectrum of an RF pulse.
Thus, we initially examine the simplest case where no filter is applied to the staircase waveform to observe the relationship between the sampling rate and gate fidelity.
In Fig.~\ref{fig:samp-rabi}, the blue line shows the simulation results of Z-errors while varying sampling time $t_s$.
When $t_s = (n \times 1)$~ns ($n=1,2,...$), a significant Z-error occurs.
Since the carrier frequency of the RF pulse is 8500~MHz, when $t_s = (n \times 1)$~ns ($n=1,2,...$),
frequency components of the RF pulse have a peak at 7500~MHz (the control qubit frequency).
We observe that frequency components at the control qubit frequency cause an oscillation between $\ket{0}$ and $\ket{1}$ of the control qubit.
For $t_s$ greater than 20~ns, Z-errors are large.
A longer $t_s$ increases frequency components other than the target qubit frequency, leading to oscillations between $\ket{0}$ and $\ket{1}$ in the control qubit.



To evaluate the deterioration of gate fidelity due to oscillations between $\ket{0}$ and $\ket{1}$, we simulate gate fidelity of CR gates using staircase waveforms with the sampling rate 1000~MHz or 1040~MHz. 
Table~\ref{tab:fidelity-samp} summarizes the results, which show that
large oscillations between $\ket{0}$ and $\ket{1}$ reduce gate fidelity.
Therefore, tuning the sampling rate of staircase waveforms to maximize gate fidelity is essential.

\begin{table}[t]
    \centering
    \caption{Z-errors and infidelity of staircase waveforms with different sampling rates.}
    \begin{tabular}{c|c|c} \hline
         Sampling rate [MHz] & 1000 & 1040  \\ \hline
         Z-error & $4.35 \times 10^{-2}$ & $3.57 \times 10^{-4}$  \\ \hline
         Infidelity [\%] & $14.9 \pm 1.39$ & $2.43 \pm 0.179$ \\ \hline 
    \end{tabular}
    \label{tab:fidelity-samp}
\end{table}


Here, by applying an LPF to staircase waveforms, we can drastically reduce the error introduced by oscillations between $\ket{0}$ and $\ket{1}$.
In Fig.~\ref{fig:samp-rabi}, the red line shows Z-errors when staircase waveform pulses are given to the control qubit through an LPF.
A first-order LPF with 100~MHz cutoff frequency is implemented.
An oscillation between $\ket{0}$ and $\ket{1}$ of control qubit ($q_1$) is reduced regardless of $t_s$ since the frequency components other than target qubit frequency are entirely suppressed by the LPF.

\begin{table}[t]
    \centering
    \caption{Z-errors in square waveforms.}
    \begin{tabular}{c|c|c|c} \hline
        & & \multicolumn{2}{|c}{Square} \\ \cline{3-4}
         & Ideal & Without LPF & With LPF\\ \hline
        Z-error & $1.56 \times 10^{-4}$ & $85.2 \times 10^{-4}$ & $0.734 \times 10^{-4}$ \\ \hline
    \end{tabular}
    \label{tab:rabi-cr-sq}
\end{table}

\subsubsection{Square waveform (AWG-less mode)}
The best pulse length $T_p$ 
is obtained by sweeping the pulse length with a fixed amplitude of 300~MHz.
The optimal pulse length is $166$~ns. 
Table~\ref{tab:rabi-cr-sq} shows Z-error results
where two
types of square waveforms are fed into the control qubit.
Although naive square waveform without an LPF exhibits a large unwanted oscillation between $\ket{0}$ and $\ket{1}$,
LPFs can effectively suppress its impact.
Table~\ref{tab:fideity-square} shows gate fidelity of two-qubit gate operations using CR gates based on the ideal or square waveform with an LPF.
The gate fidelity of the square waveform with an LPF is smaller by about 0.3~\% than that of the ideal waveform.
However, this reduction is negligibly small for the aforementioned reason.

\begin{table}[tb]
    \centering
    \caption{Randomized benchmarking results of two-qubit gates.}
    \begin{tabular}{c|c|c} \hline
         & Ideal & Square with LPF\\ \hline
        Infidelity [\%] & $1.40 \pm 0.147$  & $1.69 \pm 0.0791$ \\ \hline
    \end{tabular}
    \label{tab:fideity-square}
\end{table}

\subsection{Pulse length optimization}
\label{sec:plen}
The above discussion implies that the square waveform with an LPF can sufficiently suppress unwanted transitions.
This paper aims to achieve high gate fidelity with a low-cost circuit.
Therefore, square waveform RF pulses are used to simplify pulse generation further.
This section discusses the optimization of square waveform RF pulses to enhance gate fidelity.

\begin{figure}[t]
    \centering
    \includegraphics[clip, width=0.9\columnwidth]{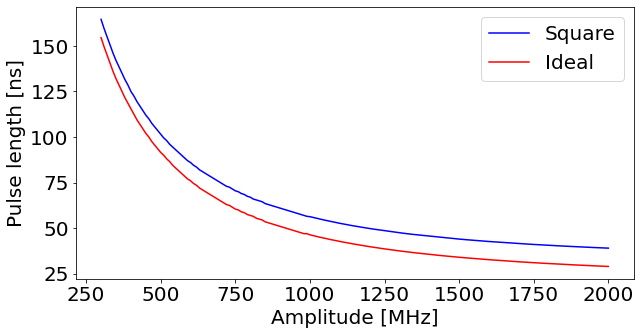}
    \caption{Optimal pulse length of CR gate while varying the amplitude.}
    \label{fig:ampl-plen}
\end{figure}

As explained in Section~\ref{sec:fidelity}, another factor determining gate fidelity is qubit decay.
A shorter RF pulse length increases gate fidelity since an RF pulse length determines the error introduced by the qubit decay.
The pulse length of the CR gate is typically longer than that of the DRAG pulse.
To minimize the effect of the qubit decay,
we shorten the pulse length of the CR gate by increasing the amplitude.
Fig.~\ref{fig:ampl-plen} shows the optimal pulse length of the ideal waveform with $T_r = 10$~ns and square waveforms (with an LPF) for CR gates while varying the amplitude from 300~MHz to 2000~MHz by 10~MHz.

\begin{figure}[t]
    \centering
    \includegraphics[clip, width=0.9\columnwidth]{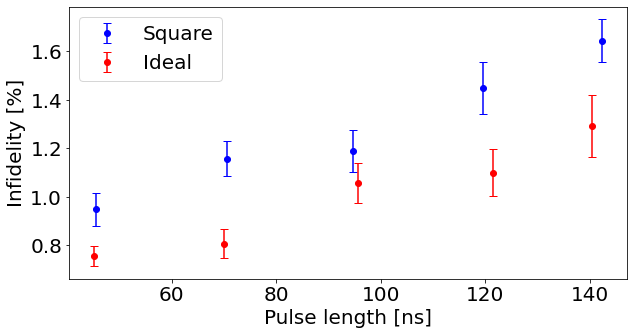}
    \caption{Randomized benchmarking results of two-qubit gates while varying the pulse length in ideal and square waveforms.}
    \label{fig:fidelity-plen}
\end{figure}

To evaluate the effect of a pulse length on gate fidelity, this paper simulates gate fidelity of two-qubit gate operations using the ideal waveform or the square waveform with an LPF.
We examined five amplitude and pulse length pairs in Fig.~\ref{fig:fidelity-plen} for both the ideal waveform and the square waveform.
It is apparent that shorter pulse lengths improve gate fidelity.
The results of a 45~ns pulse length show that the gate fidelity of square waveforms is smaller by at most 0.5\% less than that of ideal waveforms. 
Again, this reduction is negligibly small for the aforementioned reason.

\subsection{Performance comparison}
\label{sec:experiment}
This section evaluates circuit costs and gate fidelity of existing AWG-based qubit controllers and our architecture. 
The baseband sampling rate is not fast enough to generate ideal waveforms in existing AWG-based qubit controllers.
Therefore, we approximate an ideal waveform using the staircase waveform RF pulse generated by our architecture shown in Fig.~\ref{fig:proposed-controller}. 
The sampling rate of staircase waveforms is set to the baseband sampling rate of AWGs, meanwhile, our architecture adopts staircase waveforms with slow $f_s$ (low-cost AWG mode) and square waveforms (AWG-less mode).

\begin{table}[t]
    \centering
    \caption{Circuit costs and gate fidelity for two-qubit gate operations. 
    "Std.": standard deviation.}
    \begin{tabular}{l|c|c|c|c} \hline
         & & & \multicolumn{2}{|c}{Ours}  \\ \cline{4-5}
         & RT \cite{keysight} & Cryo \cite{cryo-controller-ISSCC2022} & Stair & Square \\ \hline
         Sampling rate [GS/s] & 2.4 & 1 & 0.5 & N/A \\ \hline
        \#(Waveform points) & 264 & 110 & 55 & 3 \\ \hline
        Infidelity (Mean) [\%] & $1.29 $ & $1.23 $ & 1.26 & $1.15 $ \\ \hline
        Infidelity (Std.) [\%] &  0.0630 & 0.0477 & 0.0888 & 0.0734 \\ \hline
    \end{tabular}
    \label{tab:cost-fidelity}
    \vspace{-5mm}
\end{table}

Next, we evaluate the circuit costs and the gate fidelity of two-qubit gate operations.
We target the situation
where a CR gate pulse and a DRAG pulse are fed into
the control qubit.
Table~\ref{tab:cost-fidelity} summarizes the circuit costs and gate fidelity.
A 70~ns pulse length is utilized for a CR gate pulse.
Based on the results of Fig.~\ref{fig:ampl-plen}, the corresponding amplitude is also selected.
We utilize the DRAG pulse optimized in Section~\ref{sec:DRAG}.
The pulse length is 20~ns.
The optimal amplitude is the same as the value shown in 
the column "RT" corresponds to the AWG in the qubit controller at room temperature \cite{keysight}.
The column "Cryo" corresponds to the AWG in the cryo-CMOS qubit controller \cite{cryo-controller-ISSCC2022}.
"Sampling rate" represents the baseband sampling rate of the AWGs.
"\#(Waveform points)" represents the number of waveform points for generating one CR gate pulse with a 70~ns pulse length and one DRAG pulse with a 20~ns pulse length.
Note that a DRAG pulse has two waveforms for in-phase and quadrature pulses.
This value is obtained by calculating $T_p / f_s$, where $T_p$ is a pulse length and $f_s$ is a sampling rate.
In our architecture, for example, the low-cost AWG mode (staircase waveform) uses a slow $f_s$ while maintaining comparable gate fidelity to existing approaches.
As a result, the number of waveform points is reduced by half compared with existing approaches.
We can further reduce the number of points by relaxing $f_s$.
In the AWG-less mode (square waveform) of our architecture, a pulse source generates a pulse with a single amplitude.
This drastically reduces the memory capacity since our architecture has to store only the amplitude of the square waveform.
If we count the amplitude of each pulse as one waveform point, the number of waveform points is 3 (a DRAG in-phase pulse, a DRAG quadrature pulse, and a CR gate pulse).
This means that the number of
waveform points is reduced by
two orders of magnitude compared with existing approaches.
Evaluation results of gate fidelity demonstrate employing our succinct architecture does not compromise the accuracy of quantum gate operations.

\section{Conclusion}
\label{sec:conclusion}
This paper proposed a cost-effective RF pulse generator architecture for superconducting qubits.
The proposed architecture simplifies cost-intensive AWGs for scalable implementation.
Properly tuned square waveforms can be utilized without significant gate fidelity degradation.
This paper also proposed the pulse tuning method for maximizing gate fidelity.
Evaluation results indicate that the number of sampling points, a key metric for evaluating the cost of RF pulse generators, can be reduced by two orders of magnitude compared to state-of-the-art commercial pulse generators.

\section*{Acknowledgment}
This work was supported by JST Moonshot R\&D Grant Number JPMJMS226A.

\bibliographystyle{ACM-Reference-Format}
\bibliography{qubit_control}


\end{document}